\documentclass{ws-jai}
\usepackage[flushleft]{threeparttable}
\usepackage{subfigure}
\begin{document}

\catchline{}{}{}{}{} 

\markboth{Ninan et al.}{TIRSPEC : TIFR Near Infrared Spectrometer and Imager}

\title{TIRSPEC : TIFR NEAR INFRARED SPECTROMETER AND IMAGER}

\author{J. P. Ninan$^{1*}$, D. K. Ojha$^1$, S. K. Ghosh$^{1,2}$, S. L. A. D'Costa$^1$, M. B. Naik$^1$, S. S. Poojary$^1$, P. R. Sandimani$^1$, G. S. Meshram$^1$, R. B. Jadhav$^1$, S. B. Bhagat$^1$, S. M. Gharat$^1$, C. B. Bakalkar$^1$, T. P. Prabhu$^3$, G. C. Anupama$^3$ and D. W. Toomey$^4$}

\address{
$^1$Tata Institute of Fundamental Research, Homi Bhabha Road,
Colaba, Mumbai 400 005, India; $^*$ninan@tifr.res.in\\
$^2$National Centre for Radio Astrophysics, Tata Institute of Fundamental Research, Pune 411 007, India\\
$^3$Indian Institute of Astrophysics, Koramangala, Bangalore 560 034, India\\
$^4$Mauna Kea Infrared, LLC, 21 Pookela St.Hilo, HI 96720, USA
}

\maketitle

\footnotetext[1]{$^*$Corresponding author.}

\begin{history}
\received{(to be inserted by publisher)};
\revised{(to be inserted by publisher)};
\accepted{(to be inserted by publisher)};
\end{history}

\begin{abstract}
We describe the TIFR Near Infrared Spectrometer and Imager (TIRSPEC) designed and built in collaboration with M/s. Mauna Kea Infrared LLC, Hawaii, USA, now in operation on the side port of the 2-m Himalayan Chandra Telescope (HCT), Hanle (Ladakh), India at an altitude of 4500 meters above mean sea level (amsl). The TIRSPEC provides for various modes of operation which include photometry with broad and narrow band filters, spectrometry in single order mode with long slits of 300" length and different widths, with order sorter filters in the \textit{Y}, \textit{J}, \textit{H} and \textit{K} bands and a grism as the dispersing element as well as a cross dispersed mode to give a coverage of 1.0 to 2.5 $\mu m$ at a resolving power R of $\sim$1200. The TIRSPEC uses a Teledyne 1024 x 1024 pixel Hawaii-1 PACE array detector with a cutoff wavelength of 2.5 $\mu m$ and on HCT, provides a field of view of 307" x 307" with a plate scale of 0.3"/pixel. The TIRSPEC was successfully commissioned in June 2013 and the subsequent characterization and astronomical observations are presented here. The TIRSPEC has been made available to the worldwide astronomical community for science observations from May 2014.

\end{abstract}


\section{Introduction}
Near infrared astronomy (NIR) using medium sized (2-4 m) telescopes has greatly opened up the study of many astrophysically cool objects like evolved giants and supergiants, low mass stars, red and brown dwarfs and Galactic star-forming regions. In the recent past there have been many developments in the field of NIR detector materials and arrays which have brought NIR astronomical observations almost on a par with the optical observations. Within the Indian astronomical scene, a few NIR observational facilities exist viz., Indian Institute of Astrophysics (IIA) NIRCAM on the 2-m Himalayan Chandra Telescope (HCT), Hanle (Ladakh) and the Physical Research Laboratory's (PRL) NICMOS III camera on the 1.2-m PRL telescope at Mount Abu. Recently, Tata Institute of Fundamental Research (TIFR) has also developed a NIR Camera (1.0 to 3.7 $\mu m$) \citep{ojha12b,naik12,ninan13} based on the Raytheon InSb 512 x 512 pixel Focal Plane Array (Aladdin III Quadrant) which is currently being used with the 2-m Inter-University Centre for Astronomy and Astrophysics's Girawali Observatory telescope, near Pune. Besides this, TIFR has also been observing in the far infrared (FIR) band (120 to 220 $\mu m$) using the TIFR 1-m FIR balloon borne telescope \citep{ghosh88,ghosh10}. To complement these studies in the NIR band, the need was felt for a NIR spectrometer in the 1.0 to 2.5 $\mu m$ range, which could be used with one of the 2-m class telescopes available in India. To meet this need, the TIFR Near Infrared Spectrometer and Imager (TIRSPEC) \citep{ojha12} was conceived, to provide spectrometry in the 1.0 to 2.5 $\mu m$ band with a medium spectral resolving power of $\sim$1200 to be used on one of the side ports of the 2-m HCT at Hanle (Ladakh) (4500 m amsl), along with the existing optical Himalaya Faint Object Spectrograph and Camera on the main port, to allow for near simultaneous observations in the optical and NIR bands.

TIRSPEC was developed in collaboration with M/s. Mauna Kea Infrared, LLC (MKIR), Hawaii, USA, to provide both imaging as well as spectroscopic capabilities in the 1.0 to 2.5 $\mu m$ NIR wavelength range and was designed around a Teledyne 1024 x 1024 pixel Hawaii-1 PACE array detector as the focal plane array. When mounted on the side port of the 2-m HCT at Hanle (Ladakh), the field of view (FoV) in the imaging mode is 307" x 307" with a plate scale of 0.3"/pixel, while in the spectroscopic mode it covers a wavelength range from 1.0 to 2.5 $\mu m$ with a spectral resolving power of $\sim$1200. The first light for this instrument on the HCT was observed on June 21, 2013 and after the characterization and engineering observational runs, the TIRSPEC was released to the worldwide astronomical community for science observations from May 1, 2014.

In this paper, we describe the technical details of TIRSPEC (Section \ref{sec:instrumentoverview}), and also the characterization and performance analysis done before releasing the instrument to the users for scientific observations (Section \ref{sec:characterization}). We also present the performance analysis of TIRSPEC as related to the Hanle telescope site and some samples of science observations conducted during the engineering runs (Section \ref{sec:hctperformance}). The data reduction tool developed for TIRSPEC is explained in Section \ref{sec:pipeline} and we conclude with the present status and some of the planned upgrades which are to be carried out in the near future.

\section{Instrument Overview}
\label{sec:instrumentoverview}
\subsection{Structure}

\begin{figure}[h]
\begin{center}
\includegraphics[width=0.35\textwidth]{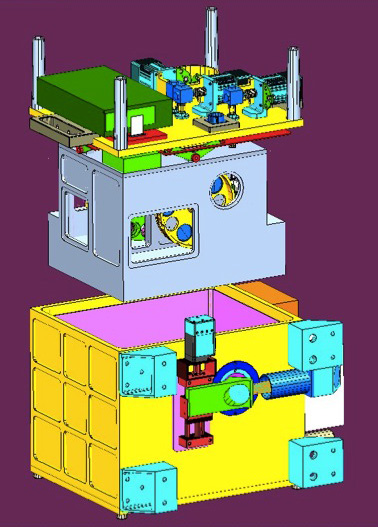} 
\end{center}
\caption{Exploded front view of the TIRSPEC cryostat. The yellow tub and top plate are the outer vacuum jacket. The pink color shields are floating radiation shields. The grey enclosure is the LN2 cooled enclosure which houses the optics.}
\label{fig:Dewarexploded}
\end{figure}

The TIRSPEC has an overall envelope size of 481 mm (width) x 600 mm (length) x 500 mm (height). The cryostat is designed as a double bath-tub which allows all the vacuum penetrations to be from the top plate of the cryostat. Figure \ref{fig:Dewarexploded} shows an exploded view of the cryostat with the outer cube providing the vacuum jacket. 
The optical bench is mounted within the liquid Nitrogen (LN2) cooled enclosure which hangs inside the outer vacuum jacket from 4 rigid fiber-glass V-trusses connected to the top plate. Thermally insulated mirror finished floating shields line the inner side of the vacuum jacket to provide good radiation shielding and to improve LN2 hold time. The LN2 tank can hold about 8.6 liters of LN2 which provides a hold time of more than 36 hours at the observatory site. The array controller electronics and the power supply for the motors and electronics are mounted on one side of the cryostat. The three motors for the slit wheel and the two filter wheels are mounted on the outer side of the top plate and are coupled to the filter and slit wheels through right angled gears and ferrofluidic vacuum feed-throughs. A calibration lamp assembly provides for spectral line calibration of the spectrometer using an argon lamp with a movable plane mirror which can be moved in and out of the optical path using an Animatics SmartMotor and linear stage arrangement. The calibration box also hosts a tungsten lamp for continuum flats with an integrating sphere.

\subsection{Optics}
\begin{figure}[h]
\begin{center}
\includegraphics[width=0.5\textwidth]{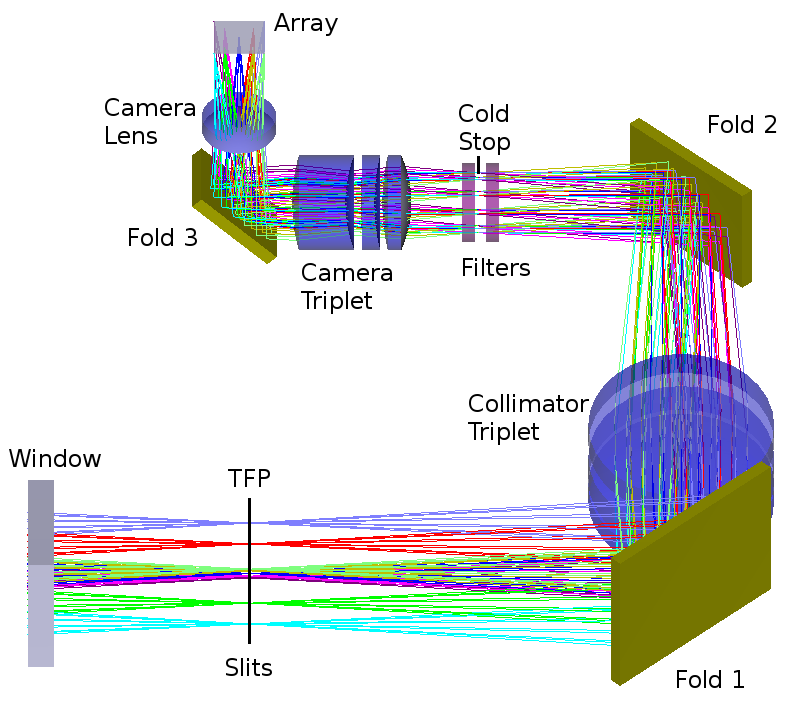} 

\end{center}
\caption{TIRSPEC's folded optics design.}
\label{fig:OpticsDesign}
\end{figure}

\begin{figure}[h]
\begin{center}
\includegraphics[width=.9\textwidth]{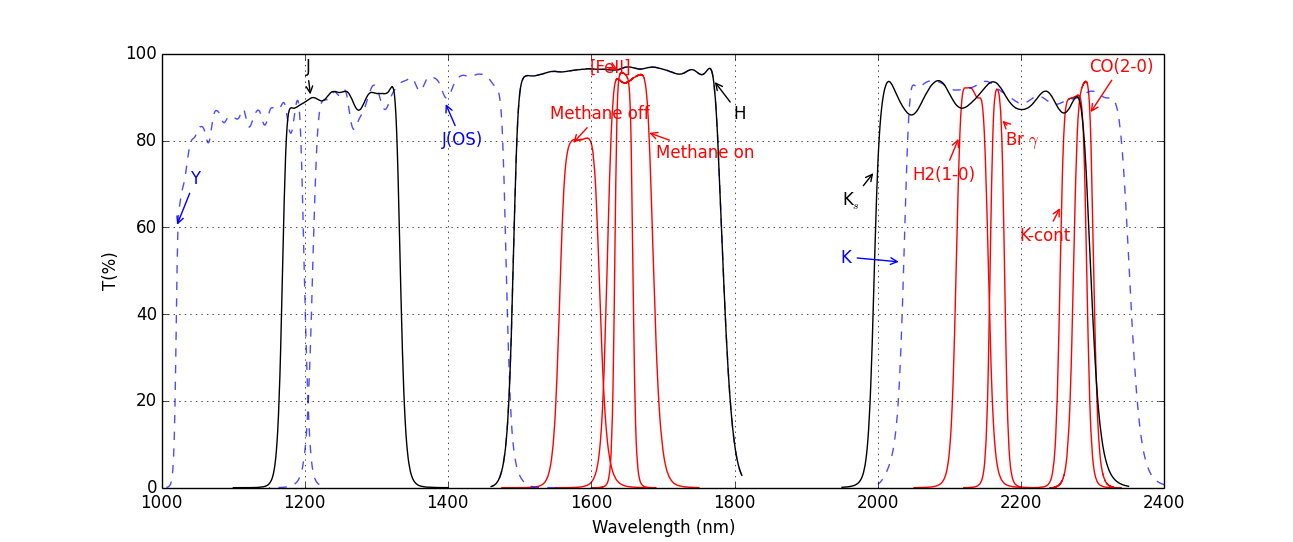} 
\end{center}
\caption{Transmittance curves of various filters on TIRSPEC. The black curves are the broad band \textit{J}, \textit{H} and \textit{K$_s$} filters, the red curves are the narrow bands, and the blue dashed curves are the order sorters for spectroscopy.}
\label{fig:filtercurves}
\end{figure}

TIRSPEC's optics were designed to be used on the side port of the 2-m HCT, to provide an image scale of 0.3"/pixel covering a FoV of 307" x 307". Figure \ref{fig:OpticsDesign} shows the layout of the optics. 
The f/9.2 beam from HCT enters the instrument via the $CaF_2$ window. At the telescope focal plane (TFP) there is a 11 position slit wheel which has 5 short slits (10" length) and 5 long slits (300" length) with slit widths of 0.9", 1.48", 1.97", 2.96" and 7.92", as well as one open position for the imaging mode. A collimator system consisting of a $BaF_2$-$LiF$-$BaF_2$ lens triplet is used to obtain a good quality achromatic image of the telescope aperture at the cold stop which is sandwiched between two 12 position filter wheels. The first filter wheel has broad band filters (\textit{J}, \textit{H}, \textit{K$_s$}) and order sorter filters (\textit{Y}, \textit{J}, \textit{H}, \textit{K}) and two cross dispersing grisms (\textit{HK} and \textit{YJ}). The second filter wheel has seven narrow band filters, one grism, block for taking dark images and an open position. Table \ref{table:filters} lists the details of imaging filters and Figure \ref{fig:filtercurves} shows the transmission curves of all the available imaging and order sorter filters.

The collimated beam is then imaged on to the focal plane array detector using a $BaF_2$-$LiF$-$ZnSe$
triplet lens and a $BaF_2$ 
singlet lens. To fit the entire optics inside the LN2 cooled enclosure, the optical path is folded at three locations using gold coated fused silica plane mirrors. 

The focal plane array used in TIRSPEC is a HAWAII-1 PACE Array\footnote{HgCdTe Astronomy Wide Area Infrared Imager - 1, Producible Alternative to CdTe for Epitaxy}, manufactured by \textit{Teledyne Scientific \& Imaging, LLC, USA}. It is a 1024 x 1024 array, with a pixel size of 18 $\mu m$ square and has a cutoff wavelength of 2.5 $\mu m$. The array has HgCdTe detector layer on top and a silicon readout layer below which contains readout amplifiers and associated circuits. Both the layers are connected by indium bumps at each pixel \citep{hodapp96,rieke07}. 
The array has four quadrants which are read out simultaneously. Read out starts from the bottom left corner pixel to the top right corner pixel of each quadrant at a rate of 3 $\mu s$/pixel. The detector is operated at a temperature of 76-77 degrees K at Hanle (Ladakh).

\begin{wstable}[h]
\caption{Imaging filters.}
\begin{tabular}{@{}cc@{}} \toprule
Filter & Band pass ($\mu m$)\\
\colrule
\textit{J} & 1.17 - 1.33 \\
\textit{H} & 1.49 - 1.78 \\
\textit{K$_s$} & 1.99 - 2.30 \\
\colrule
Methane off & 1.584 (3.6 \%) \\
$[$Fe II$]$ & 1.645 (1.6 \%) \\
Methane on & 1.654 (4.0 \%) \\
H2 (1-0) S(1)& 2.1239 (2.0 \%) \\
Br $\gamma$ & 2.166 (0.98 \%) \\
K-Cont & 2.273 (1.73 \%) \\
CO (2-0) & 2.287 (1.33 \%) \\
\botrule
\end{tabular}
\label{table:filters}
\end{wstable}

Spectropscopy using TIRSPEC can be done using the single order mode or the cross dispersed mode. In both modes, a \textit{YJHK} grism is positioned in the second filter wheel. This grism has a groove width of 12.4 $\mu m$ and a prism angle of 21.9$^o$ which disperses \textit{Y}, \textit{J}, \textit{H} and \textit{K} in its 6th, 5th, 4th and 3rd orders respectively. To obtain the spectrum of each order separately, in the single order mode, the corresponding order sorter filter is positioned in the first filter wheel which blocks the other orders.

In the cross dispersed mode, we can simultaneously obtain spectra in two orders. Two modes are provided which cover the \textit{Y} and \textit{J} windows or the \textit{H} and \textit{K} windows. This is accomplished by positioning the \textit{YJHK} grism in one filter wheel and a cross dispersing grism in the second wheel. The cross dispersing grism is mounted 90 degrees rotated with respect to the \textit{YJHK} grism and has a custom order sorter filter that covers both orders. The first order dispersion of the \textit{YJ} cross dispersed grism with groove width of 6.2 $\mu m$ and prism angle of 8$^o$ is used to obtain cross dispersed \textit{Y} and \textit{J} spectrum. Similarly, the first order dispersion  of the \textit{HK} cross dispersed grism  with groove width of 15.4 $\mu m$ and prism angle of 5$^o$ is used to obtain cross dispersed \textit{H} and \textit{K} spectrum. All the grisms are directly ruled grisms. Table \ref{table:spectrumrange} shows the  wavelengths covered in each mode.

\begin{wstable}[h]
\caption{Wavelength coverage in different spectroscopy modes.}
\begin{tabular}{@{}cc@{}} \toprule
Spectroscopy Order & Wavelength Range ($\mu m$)\\
\colrule
\textit{Y} & 1.02 - 1.20 \\
\textit{J} & 1.21 - 1.48 \\
\textit{H} & 1.49 - 1.78 \\
\textit{K} & 2.04 - 2.35 \\
\colrule
\textit{YJ}\tnote{*} & 1.02 - 1.49 \\
\textit{HK}\tnote{*} & 1.50 - 1.84 \& 1.95 - 2.45 \\
\botrule
\end{tabular}
\begin{tablenotes}
\item[*] \textit{YJ} and \textit{HK} are cross dispersed modes.
\item Note: 1.84 to 1.95 $\mu m$ is not part of the spectrum in \textit{HK} due to their dispersion being outside the physical dimensions of the detector array.
\end{tablenotes}
\label{table:spectrumrange}
\end{wstable}

\subsection{Electronics and Control systems}
The HAWAII-1 PACE array is controlled by an \textit{ARC Gen III array controller} (also known as Leach Controller or SDSU Controller) from \textit{Astronomical Research Cameras, Inc. (ARC), San Diego, USA}. The array is mounted on a Fanout Board inside the cryostat. The image data is sent to the \textit{ARC-46: 8-Channel IR Video Processor} in the array controller box mounted at the side of TIRSPEC cryostat where it is amplified and digitized. The \textit{ARC-32 CCD and IR Clock Driver Board} provides the clocks to the array and the controller boards. The \textit{ARC-22: 250 MHz Fiber Optic Timing Board, Gen III} generates the waveforms as well as communicates with a PCI (Peripheral Component Interconnect) interface board, the \textit{ARC-64: 250 MHz PCI Interface board, Gen III} mounted in the TIRSPEC control computer, via a duplex fiber optic cable link which transfers data and commands between the computer and the array controller box. 
The cryostat and array temperatures are monitored using a \textit{Lakeshore temperature controller Model 335}.

The slit, filter wheels and calibration box mirror are moved using independent SmartMotors manufactured by \textit{Animatics Corporation, Milpitas, USA}. The motors are  controlled through a standard Ethernet interface.
The slit and filter wheels have a spring loaded detent mechanism with a thin film Hall sensor to detect the position of the wheel.

\subsection{Software}
\begin{figure}[h]
\begin{center}
\includegraphics[width=0.5\textwidth]{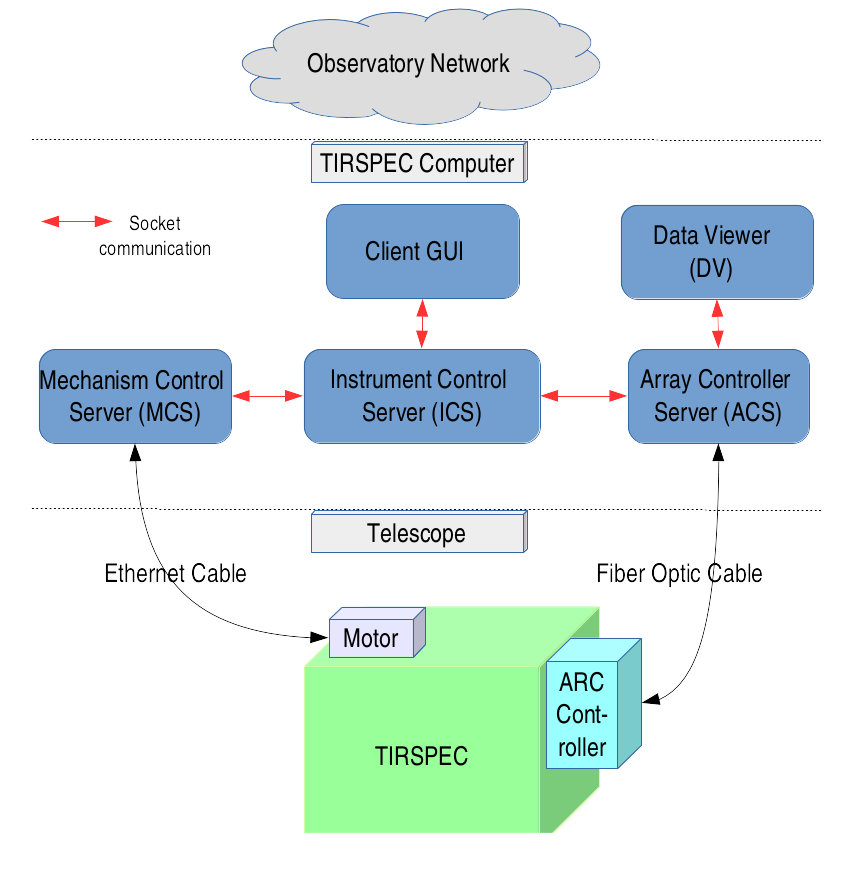} 
\end{center}
\caption{Overall structure of TIRSPEC instrument control software}
\label{fig:ControlSoftware}
\end{figure}

The TIRSPEC instrument is controlled from the TIRSPEC control computer using the main control software. The TIRSPEC control computer is connected to two independent networks with one being used to communicate with the observatory network for remote access, and the other being used to communicate with the TIRSPEC components which includes the Animatics SmartMotors and the TIRSPEC power control switches. The overall structure of the instrument control software is shown in Figure \ref{fig:ControlSoftware}. 

The Instrument Control Server (ICS) software is the central server which controls all the other servers. It communicates with the other servers via UNIX style sockets. The ICS receives commands from the client graphical user interface (GUI) as well as returns instrument status to the GUI. When the client requests a particular filter to be in position, the ICS receives the command from the GUI and if it is a valid command, it sends the instruction to the Mechanism Control Server (MCS). The MCS talks via the Ethernet link with the corresponding Animatics SmartMotor which turns the filter wheel to the desired position. When the client requests an exposure of a specified integration time, the ICS receives the command, and if the system is ready, it sends the instruction to the Array Controller Server (ACS). The ACS talks via the ARC application programming interface (API) to the ARC controller. The ARC device drivers and API are highly customized versions of the standard code provided by \textit{Astronomical Research Cameras, Inc}. Once the CPU receives an interrupt from the PCI board after the data is fully written to the system memory, ACS generates FITS\footnote{Flexible Image Transport System} image from the raw data and saves it to the hard-disk in standard FITS format. It also sends the data via a socket to the quick look viewer software DV\footnote{Data Viewer by  NASA Infrared Telescope Facility (IRTF), Hawaii, USA}. DV has tools to do basic image arithmetic as well as quick measurements to check image quality.

\subsubsection{Readout mode and image generation}
\label{sec:ReadoutAndImgGen}
When an exposure is not being taken, the array is continuously reset globally. At the beginning of any exposure, the reset pulse is stopped and the detector is read out non-destructively. The detector can be read out in various modes. For optimal results, the sample-up-the-ramp (SUTR) readout mode has been chosen as the default mode. The pixels are read at a rate of 3 $\mu s$ per pixel and since all four quadrants of the detector are read out simultaneously, it takes 0.9 seconds to complete one full read out of the array. For an integration time \textit{I}, the actual exposure time is automatically adjusted to the nearest multiple of 0.9 second and the actual integration time used is written to the header of the data file. A total of \textit{I}/0.9 non-destructive readouts of the cumulative counts in the array pixels are executed during the duration of the exposure. Thus, the raw data for each exposure is a data cube with the pixels on the two axes and time on the third axis. The flux in each pixel is then calculated by fitting a slope by linear regression to this SUTR readout data cube along the time axis. Due to the reset anomaly discussed later in Section \ref{sec:ResetAnomaly}, one needs to subtract the dark readout cube from the image data cube before linear slope fitting. The SUTR readout technique gives us unique capabilities to recover data from cosmic ray (CR) hits and saturation.
\subsubsection{Cosmic ray hit healing}
\begin{figure}[h]
\begin{center}
\includegraphics[width=0.5\textwidth]{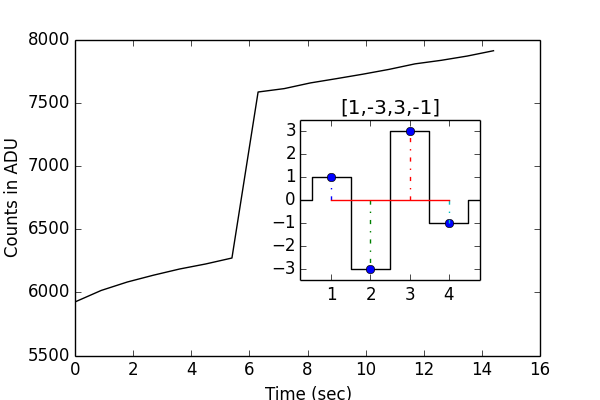} 
\end{center}
\caption{CR hit between the 5$^{th}$ and 6$^{th}$ seconds during an SUTR readout exposure. The inset shows the digital filter used to detect these events.}
\label{fig:CRhit}
\end{figure}

HCT at Hanle (4500 m amsl) is prone to much higher CR hits than lower altitude observatories. Since the SUTR readout has time resolved data, the readout counts before and after a CR hit can be used to retrieve the slope of the flux at that pixel. Figure \ref{fig:CRhit} shows the typical SUTR readout generally seen for a CR hit. The sudden jump in the ramp can be detected using a digital filter shown in inset of Figure \ref{fig:CRhit}. This filter is created by convolution of a first  difference filter and a Mexican hat filter. After identifying the CR hit location, the slope in linear sections of the SUTR data is estimated and averaged (weighted by the error in slope of each section).

\subsubsection{Saturation healing}
Another advantage in using the SUTR readout is that saturated pixels can be selectively discarded. This is especially useful when observing fields which have both bright and faint stars. The flux is estimated using only the readouts during the linear regime of the pixels containing bright saturating stars, while for the remaining pixels, the entire data is used. This effectively increases the dynamic range of the detector beyond the actual well depth.
\subsubsection{Non-linearity correction}
From SUTR readout data we have estimated the non-linearity correction coefficients for each pixel, which can be applied to reconstruct the data beyond 1\% non-linearity. This has not been done by default to any of the data. In general, all the data beyond the linear regime is discarded before fitting the slope.

\section{Characterization and Performance}
\label{sec:characterization}
\subsection{Dark current}
Dark current noise determines the sensitivity limit we can attain in the spectroscopy mode. Unlike the imaging mode, the noise in a faint star continuum signal is dominated by dark current and effective readout noise.
Dark current is measured by reading out the array cooled to 77 K (normal operating temperature) with the cold block filter in the second filter wheel. 

The measured dark current has an exponential fall off at the beginning of the exposure.
This high value at the beginning is due to the reset anomaly of the detector and is not due to high dark current. The reset anomaly is discussed in more detail in Section \ref{sec:ResetAnomaly}. Hence, the actual dark current of the instrument is estimated only after the detector output has stabilized. A straight line fit to the readout values after skipping the first 27 sec of data from the start of the exposure up to 100 sec gave a dark current value of 0.032 ADU/sec. This multiplied by gain (see Section \ref{sec:Gain}) gives us an estimate of the dark current $\sim$ 0.19 e$^{-}$/sec. This value matches with estimates of dark current by others on similar HAWAII-1 PACE 1k x 1k Array \citep{pulido03}. Stability of dark current over a time scale of one night was tested to decide the number of dark frame readouts needed to be taken during each night. It was found that it would be enough to take a few sets of dark frame readouts at the beginning and end of each night. 

\subsection{Reset Anomaly}
\label{sec:ResetAnomaly}
\begin{figure}[h]
\begin{center}
\includegraphics[width=0.45\textwidth]{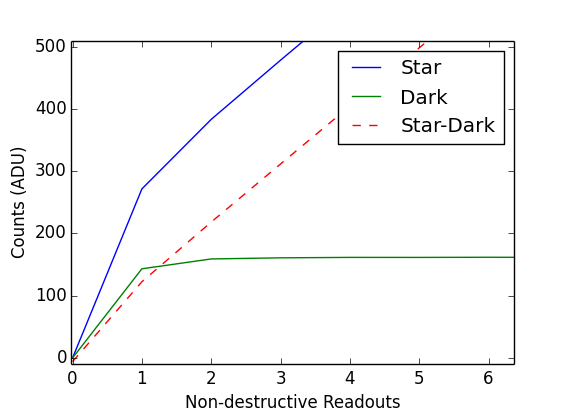} 
\end{center}
\caption{The linear slope in sample-up-the-ramp (SUTR) readout. Blue curve shows the readout of a pixel exposed to star and green line shows the dark readout. Red dashed line is the linear readout which is recovered after subtracting the non-linear dark in image data cube.}
\label{fig:FluxMinusDark}
\end{figure}

\begin{figure}[h]
\begin{center}
\includegraphics[width=0.5\textwidth]{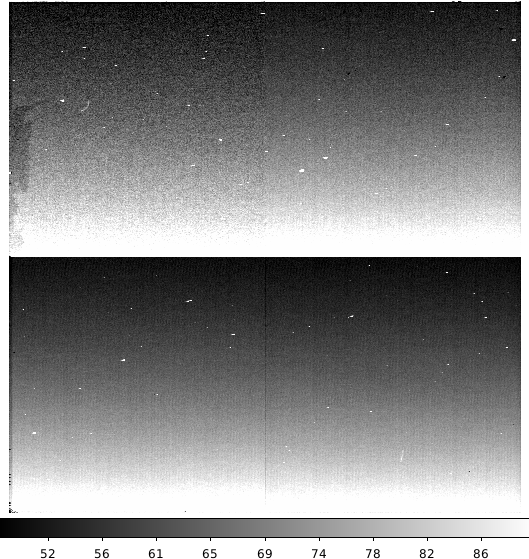} 
\end{center}
\caption{The vertical gradient in each quadrant of the raw image due to the reset anomaly in a 4 sec dark exposure. This gradient in the raw image can be removed fully by dark subtraction.}
\label{fig:rawimg}
\end{figure}

HAWAII-1 PACE array has been known to have a non-linear behaviour immediately after the detector is reset \citep{finger00,pulido03}. In spite of resetting at the same frequency as the readout, to reduce the effects of the reset anomaly as suggested by \citet{riopel04}, we still see the reset anomaly in our readouts. The green curve in Figure \ref{fig:FluxMinusDark} shows the non-linear output obtained in the dark readout due to the reset anomaly. 
Since this is a strong function of the time that has elapsed since the reset at the beginning of exposure, we get a vertical gradient in the raw image frames in each of the quadrants as shown in Figure \ref{fig:rawimg}.
One solution to this anomaly is to take long exposures and discard the first few readouts. But, this cannot be done for imaging in the \textit{H} and \textit{K} bands or for bright stars due to saturation. Hence, it is crucial to decouple this non-linear readout portion to obtain the true linear slope corresponding to the flux.
This non-linear behavior was found to be an additive effect, and can be subtracted out to obtain the linear curve corresponding to the actual flux falling on the detector. Figure \ref{fig:FluxMinusDark} shows the linear curve (dashed line) recovered after subtracting the dark readout (green curve). \citet{finger00} had also shown this additive nature of the reset anomaly.

\begin{figure}[h]
\begin{center}
\includegraphics[width=0.5\textwidth]{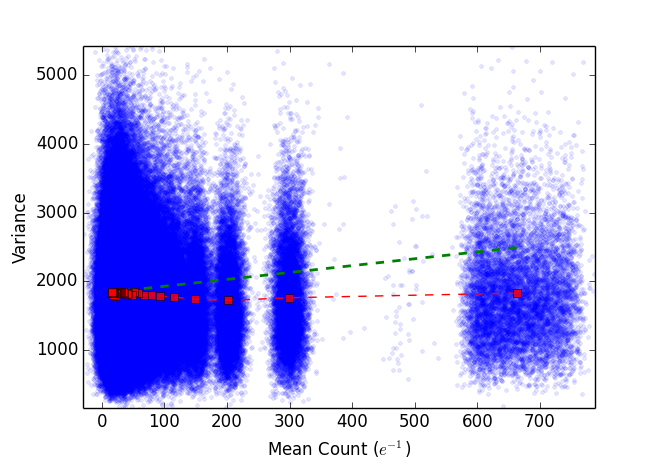} 
\end{center}
\caption{Variance versus mean counts. The green dashed line shows the expected increase if the counts are of Poissonian statistics like dark current, and the red squares represent the measured variance obtained by averaging of the blue points, each representing pixels in a 100 x 100 sub-array with origin at (400,400) in the full array.}
\label{fig:CountVsVariance}
\end{figure}

To study the noise statistics of these large counts obtained at the initial seconds of the exposure, multiple dark SUTR readouts were taken and the dark current versus the variance were plotted. If the large counts obtained in the initial seconds of the exposure are due to dark current, we expect a Poisson noise variance in them. Figure \ref{fig:CountVsVariance} shows the counts versus the variance. Blue dots are points corresponding to pixels from a good 100x100 subarray of pixels with origin at (400,400) of the full array. The red squares correspond to the mean of the blue points corresponding to each set of readouts. The plot shows the red points are almost constant and do not increase linearly with the counts. The green dashed line shows the expected increase in variance if the counts had Poissonian statistics, as a function of the counts. This implies that the reset anomaly does not have any significant contribution in the final noise of the data if we just discard the first few readouts (see Section \ref{sec:ReadoutAndImgGen}).

\subsection{Readout Noise}
\begin{figure}[h]
\begin{center}
\includegraphics[width=0.5\textwidth]{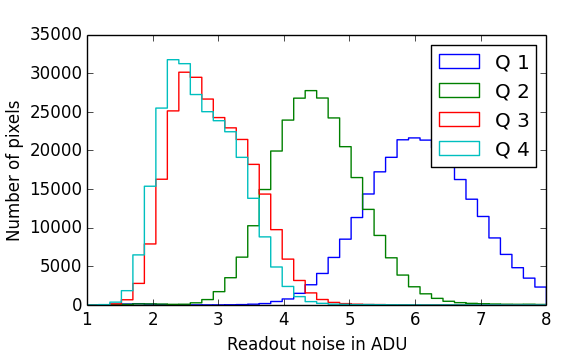} 
\end{center}
\caption{Histogram of readout noise for single read out in the array. The four histograms correspond to the four quadrants in the array.}
\label{fig:readnoise}
\end{figure}

Readout noise mainly comes from the amplifier noise in the electronics circuit. These were found to be uncorrelated in our 0.9 sec readout time interval of each pixel during an exposure. Since dark current was found to be very small compared to the readout noise after 27 sec of integration, we used difference of consecutive frames in SUTR readout of a dark frame, to estimate $\sqrt{2}$ times readout noise. The readout noise was found to be different for each quadrant. Figure \ref{fig:readnoise} shows the histogram of readout noise for single readout of different pixels in the array. 
The mean readout noise for single readout for each quadrant is 6.1, 4.5, 2.9 and 2.8 ADU, with an average value of  4.1 ADU (i.e. $\sim$ 25 e$^{-}$) for the full array. It should be noted that this value is the readout noise for a single readout, hence readout noise contribution in the flux estimate is usually much less than this, and it is a function of the number of readouts in a single exposure and flux. Detailed treatment of error contribution of readout noise in flux calculation in SUTR (line-fitting) mode is available in \citet{robberto07}.

\subsection{Gain}
\label{sec:Gain}
\begin{figure}[h]
\begin{center}
\includegraphics[width=0.5\textwidth]{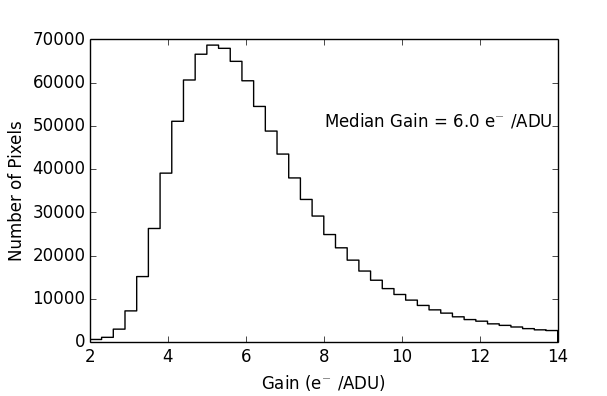} 
\end{center}
\caption{The histogram of ADU to e$^-$ gain of each pixel in the array.}
\label{fig:Gainhist}
\end{figure}

Gain is the multiplicative factor which converts ADU counts obtained from the detector array to electrons. This is crucial to estimate the actual photon counts obtained to estimate photon noise. For measuring the gain, we took multiple exposures of different exposure times with the array exposed to the telescope enclosure with the \textit{K} filter. Since the thermal flux from the telescope is stable over the short duration of our exposures, these images are purely photon noise limited. For each pixel we calculated the slope of ADU flux counts versus variance. Since photon noise follows Poisson statistics, actual electron counts versus variance  will have a unity slope. Hence, the multiplicative gain factor was calculated by taking the inverse of the slope in ADU counts versus variance plot. Figure \ref{fig:Gainhist} shows the histogram of the gain obtained for each pixel in the array. The median gain of the detector was found to be $\sim$ 6 $e^{-}$/ADU.

\subsection{Saturation level (Linear range)}
Unlike optical CCDs, NIR detectors have strong non-linearity near the saturation level of the pixels. 
 Each pixel saturates and deviates from the linear regime at slightly different counts. Using the SUTR readout data, we estimated the upper limit of linear regime for each pixel separately. The median saturation point of the pixels was found to be at $\sim$16000 ADU, subtracting the median bias level of $\sim$5700 ADU, gives us an effective well depth of $\sim 10300$ ADU ($\sim 61800 e^{-}$). By default, all the data in the non-linear regime are discarded, but we have added a provision in the data reduction software for users to do non-linearity correction in the non-linear regime and save the corrected data if needed (see Sections \ref{sec:ReadoutAndImgGen} and \ref{sec:pipeline}).

\subsection{Bad, Hot and Cold pixels}
Bad pixels are defined to be those pixels in the array which deviate more than $8\sigma$ from the median value in the image obtained by dividing two flats taken in high and low incident flux. This set includes all the pixels whose gain is varying with time or are completely dead. 
In our detector array, most of such pixels are along the edges. The percentage of the bad pixels in the array is 0.1\%. Apart from this, during some of the nights, the vertical single strip of pixels at column position 512 (at the border of quadrants) was also found to have a varying gain. Hence, we avoid objects falling on that column during our observations.

Hot pixels are defined to be pixels which have counts more than $8\sigma$ above the median value in a dark readout of 100 sec. Their fraction was found to be 2.3\% (most of them are located at the four quadrant corners heated by the glow from the nearby readout transistor). Similarly, cold pixels (pixels which have counts less than $8\sigma$ below median value in dark) fraction was found to be 0.07\%. It is to be noted that, only a subset of cold and hot pixels are actually bad pixels, which need to be masked, and the remaining can be corrected by flat fielding and dark subtraction.  For robustness of all the above calculations, standard deviation $\sigma$ was estimated from the median absolute deviation.

\subsection{Spectral resolving power}
\begin{figure}[h]
\begin{center}
\includegraphics[width=0.7\textwidth]{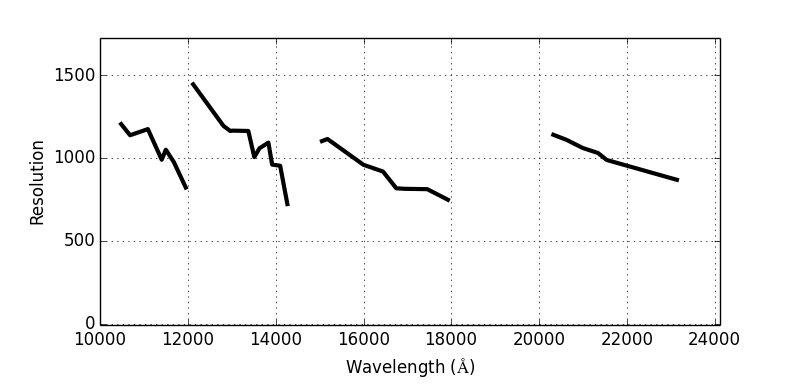} 
\end{center}
\caption{Median spectral resolving power measured as a function of wavelength from argon lamp spectrum.}
\label{fig:spectralResolution}
\end{figure}

The reciprocal linear dispersion of TIRSPEC in spectroscopy mode is $\sim$4.7 $\AA$/pixel.
The effective spectral resolving power of the spectra in each order was estimated by finding full-width-half-maximum (FWHM) of argon lamp lines through the narrow slit of 0.9" width. Figure \ref{fig:spectralResolution} shows the median $\lambda/\Delta\lambda$ spectral resolving power along the slit obtained from argon lines as a function of wavelength.

\subsection{Fringe issue}
\begin{figure}[h]
\begin{center}
\includegraphics[width=0.8\textwidth]{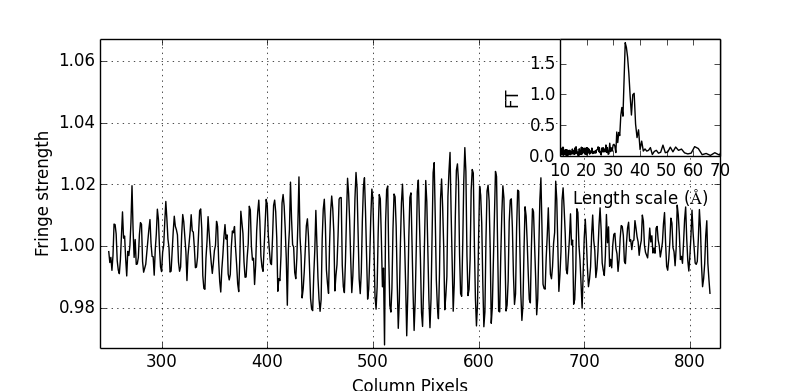}
\end{center}
\caption{Normalized fringe pattern seen in \textit{K} band spectrum. Fourier transform of the fringe pattern (shown in inset) gave the spatial size of fringes to be $\sim$36 $\AA$.}
\label{fig:fringes}
\end{figure}

Spectra in \textit{K} band show fringes in the detector. HAWAII-1 PACE arrays are known to have this issue due to interference in the sapphire substrate on which the HgCdTe detector material is grown \citep{hodapp96}. Figure \ref{fig:fringes} shows the normalized strength of fringes seen in \textit{K} band of 0.9" slit spectrum. 
The fringe pattern was found to be stable and its Fourier transform gave a typical wavelength scale equivalent to $\sim$ 36 $\AA$ in dispersed spectrum. Since this is close to our typical stellar line widths, it is not safe to use any Fourier transform based fringe removal technique. However, since the fringes are stable, it can be removed by normalizing with contemporaneous flat and standard star spectra taken  at the same position along the slit.

\subsection{Persistence issue}
HAWAII-I PACE array detectors are known to have strong persistence and this leaves a slowly decaying persistence image if exposed to very bright light. The time scale for decay of persistence image is a function of the strength of bright light which caused persistence. A shift of 23 pixels perpendicular to the dispersion direction in the position of the spectrum by the grism in TIRSPEC avoids the spectrum from overlapping on the persistence of bright star image taken for centering the star in the slit.

\section{Performance on HCT}
\label{sec:hctperformance}
\subsection{Plate Scale and Field of View}
We matched 2MASS coordinates of stars in the selected NIR standard star fields catalog \citep{hunt98} to the pixel coordinates in TIRSPEC image. The WCS submodule in Astropy\footnote{\citet{astropy13}} and optimize submodule in Scipy\footnote{Scipy is a standard package for scientific computing with Python \citep{oliphant07}} was used to fit the world coordinate parameters by least square minimization. The plate scale of TIRSPEC images obtained is 0.30 arcsec/pixel.
With the measured plate scale, we obtain a FoV of 307 $\times$ 307 arcsec$^{2}$ ($\sim$ 5 $\times$ 5 arcmin$^{2}$) for the 1024 $\times$ 1024 array.

\subsection{Sky background}
In NIR, where the sky is almost 800 times brighter than in the optical, the faint magnitude limit in the imaging mode is limited by photon noise from the sky background. To estimate the sky brightness at HCT, we observed NIR standard star fields from \citet{hunt98} on various nights. The sky brightness in \textit{H} and \textit{K} bands was found to vary depending on the humidity during the nights. This is expected, since the major contribution of flux in the \textit{H} band comes from water and OH telluric lines. The median values of sky brightness obtained in good night conditions are the following:  \textit{J} = 16.14 mag/arcsec$^{2}$, \textit{H} = 13.70 mag/arcsec$^{2}$, \textit{K$_{s}$} = 13.35 mag/arcsec$^{2}$ and \textit{K} = 13.20 mag/arcsec$^{2}$. These are comparable to other good sites like Mauna Kea (4145 m amsl) and Paranal (2635 m amsl).

\begin{figure}[h]
\begin{center}
\includegraphics[width=0.5\textwidth]{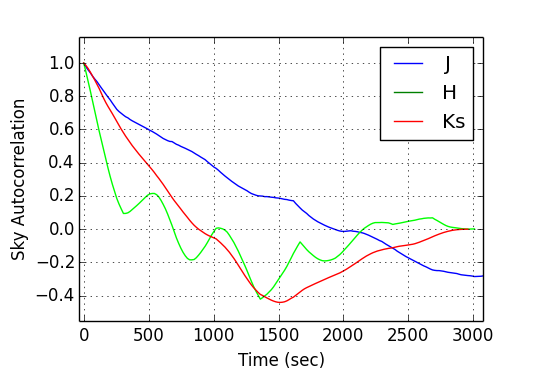} 
\end{center}
\caption{Blue, green and red curves show the auto-correlation of the \textit{J}, \textit{H} and \textit{K$_s$} sky brightness respectively. The  e-folding time scale gives the typical time scale in which sky brightness varies at the HCT site.}
\label{fig:skyautocorr}
\end{figure}

The other important related parameter is the time scale of variation in sky brightness. This determines how fast one should observe separate sky frames, to subtract from object frames (especially while observing diffuse/extended emission regions). Time scales also vary considerably from night to night. To obtain typical time scales, we monitored the sky brightness of one region continuously in intervals of 17 seconds. Figure \ref{fig:skyautocorr} shows the auto-correlation of sky brightness in the \textit{J}, \textit{H} and \textit{K$_{s}$} filters. The  e-folding time scale in auto-correlation is $\sim$600 sec in the \textit{H} and \textit{K$_{s}$} bands at HCT site, whereas the time scale is longer  in the \textit{J} band.

\subsection{Throughput of the entire system}
The throughput of the entire system was measured by counting the total flux from standard stars over an area of 4 times the FWHM of their profile. The image mode throughput percentage is 16 $\pm$ 5 \%, 20 $\pm$ 5\% and 20 $\pm$ 5\% for the \textit{J}, \textit{H} and \textit{K$_{s}$} bands respectively. This  percentage corresponds to the net throughput and it includes the atmosphere, telescope mirrors, instrument optics and quantum efficiency (QE) of the detector. These values are consistent with our initial theoretical estimate of throughput. Spectral throughput is difficult to measure by this method, because of the difficulty in placing the star at the exact center of the slit every time. However, we have a theoretical estimate of the spectral throughput to be 30\%, which does not include the QE of the detector and the slit losses.

\begin{figure}[h]
\begin{center}
\includegraphics[width=0.5\textwidth]{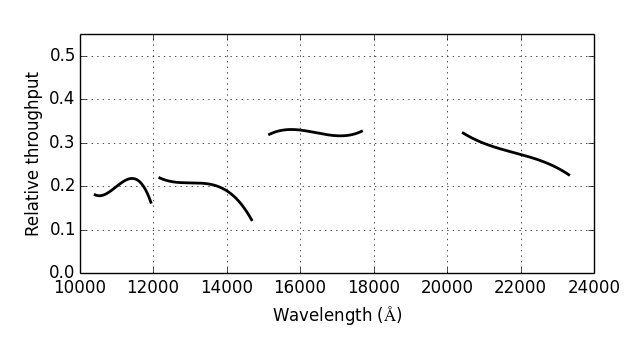} 
\end{center}
\caption{Four curves showing throughput in the cross dispersed mode normalised with respect to the corresponding single order mode in each band. This fractional reduction in net throughput is due to the additional grism used in the cross dispersed mode.}
\label{fig:crossthroughput}
\end{figure}

While the cross dispersed mode of spectroscopy has the huge advantage of simultaneously taking two orders of spectrum and even slightly extended wavelength coverage at the 2.4 $\mu m$ end, there is a trade off in the net throughput between the single order and the cross dispersed modes due to the additional grism. Figure \ref{fig:crossthroughput} shows the measured fractional throughput of the cross dispersed mode relative to the single order mode. This was obtained by taking the ratio of the continuum tungsten lamp spectra in the single and cross dispersed modes through the same slit.

\subsection{Photometric sensitivity}
\begin{figure}[h]
\begin{center}
\includegraphics[width=0.7\textwidth]{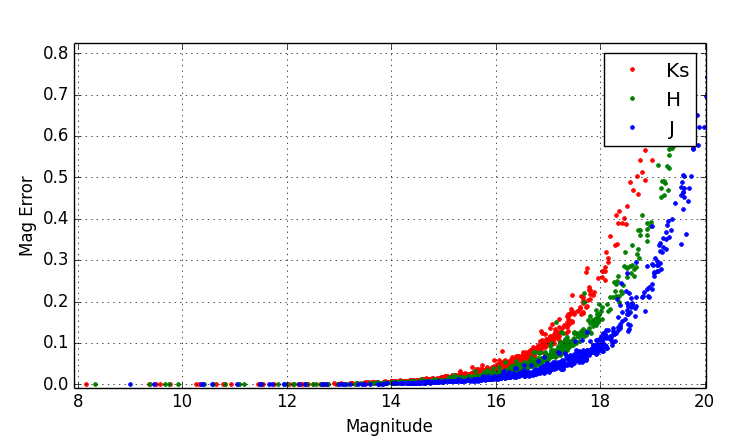} 
\end{center}
\caption{\textit{J}, \textit{H} and \textit{K$_s$} magnitudes versus magnitude errors plot for the sources in the AS 40 field, taken with 170 sec of total exposure. The photometry was carried out with an aperture of radius 1 FWHM.}
\label{fig:photoAS40}
\end{figure}

The estimate of the limiting magnitude depends on the seeing conditions. During our instrument calibration nights, FWHM of the star profile was never below 1.4" in the \textit{K$_s$} band.  To estimate the photometric sensitivity, we observed crowded standard star fields \citep{hunt98} and obtained the estimated magnitude versus  magnitude error plot. Figure \ref{fig:photoAS40} shows the \textit{J}, \textit{H} and \textit{K$_s$} magnitudes versus their magnitude errors obtained for the sources in the AS 40 field \citep{hunt98} for 170 sec of exposure at 40$^o$ elevation on a night with seeing of 1.8". The photometry was carried out with an aperture of radius 1 FWHM.

\begin{figure}[h]
\begin{center}
\includegraphics[width=0.5\textwidth]{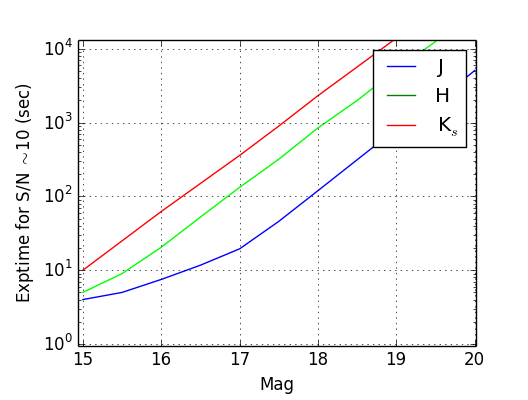} 
\end{center}
\caption{Exposure time required for 0.1 mag error (10\%) photometry versus star magnitudes in the \textit{J}, \textit{H} and \textit{K$_s$} bands.}
\label{fig:photoexptime}
\end{figure}

Figure \ref{fig:photoexptime} shows the exposure time required for 0.1 mag error (10\%) photometry for various star magnitudes on a typical night. It is generated by calculating the signal-to-noise (S/N) ratio for slope fitting \citep{robberto07} and integrating the entire S/N ratio contribution for a given Gaussian star profile of FWHM=1.6" and photometric aperture of radius 1 FWHM.

\subsection{Spectroscopic sensitivity}
\begin{figure}[h]
\begin{center}
\includegraphics[width=0.5\textwidth]{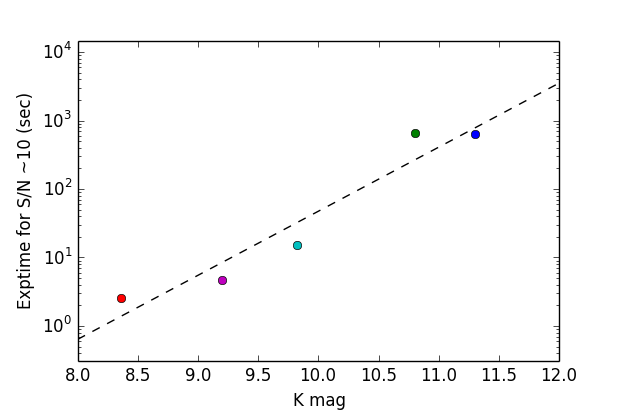} 
\end{center}
\caption{Various estimates of exposure time required in single order mode for S/N $\sim$10 in continuum in the \textit{K} band versus \textit{K} band magnitude of the source on typical nights observed from HCT site.}
\label{fig:specsensitivity}
\end{figure}

Attainable spectroscopic sensitivity for different magnitudes were estimated by taking spectra of faint stars. Figure \ref{fig:specsensitivity} shows various estimates of exposure time required for a continuum S/N $\sim$ 10 in single order mode. It is quite sensitive to the FWHM of the star profile as well as the accuracy in centering of the star inside the slit. We have found that spectra of stars upto 12  mag in the \textit{J}, \textit{H} and \textit{K} bands can be obtained under typical (1.6" seeing) night conditions.

\subsection{Color equations}
For accurate (1\%) broad band photometry, we need to correct for the filter response with stars of different color. Coefficients of 1$^{st}$ order color equations were obtained by fitting a straight line to magnitude zero points of standard stars with a wide range of colors. To establish the color equations, we have observed the color ranges from 0.45 to 2.5 for \textit{J}-\textit{K$_s$}, 0.35 to 1.76 for \textit{J}-\textit{H}, and 0.06 to 2.83 for \textit{H}-\textit{K$_s$}. To improve the statistics, we simultaneously fitted the slopes of many standard star fields, letting the airmass correction term free to vary from one field to another. Standard star fields were taken from \citet{hunt98} and red standard stars from \citet{persson98}. The color correction equations obtained for TIRSPEC at HCT site for calibrating to 2MASS magnitudes are the following.\\
\begin{equation}
\begin{aligned}
&J-K &= &  0.97\pm0.03 *(j-k)&+\beta_{JK}\\
&K-k &= &  0.11\pm0.03 *(J-K)&+\beta_{Kk}\\
&J-H &= &  1.00\pm0.05 *(j-h)&+\beta_{JH}\\
&J-j &= &  0.12\pm0.04*(J-H)&+\beta_{Jj}\\
&H-K &= &  0.82\pm0.07*(h-k)&+\beta_{HK}\\
&K-k &= &  0.07\pm0.03*(H-K)&+\beta_{Kk}\\
\end{aligned}
\label{eqns:coloreqns}
\end{equation}

\subsection{Field of View distortions} 
\begin{figure}[h]
\begin{center}
\includegraphics[width=0.5\textwidth]{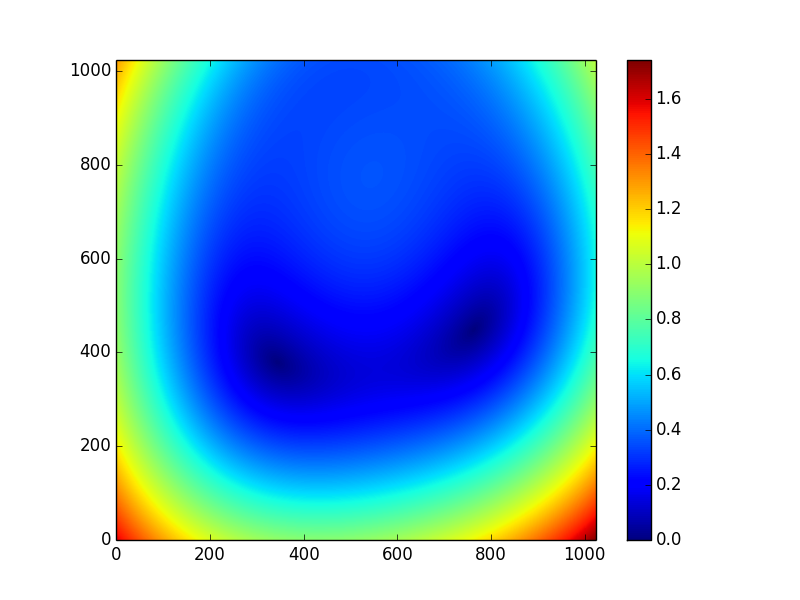} 
\end{center}
\caption{The colour map shows the magnitude of deviation from the linear estimate of FoV in units of arcsec. }
\label{fig:FoVDist}
\end{figure}

To measure any distortions in the FoV, we used the coordinates of 2MASS sources in a crowded field observed with TIRSPEC. After matching the coordinates of stars in the TIRSPEC image with the  2MASS field, we used the geomap tool in IRAF\footnote{IRAF is distributed by the National Optical Astronomy Observatory, which
is operated by the Association of Universities for Research in Astronomy, Inc., under cooperative agreement with the National Science Foundation.} to obtain second order transformation coefficients to quantify FoV distortion. A third degree general polynomial fit was also done to study the FoV distortions. Figure \ref{fig:FoVDist} shows the measured deviation from linear FoV in the array plane. The units are in arcsec. Significant FoV distortions are found only towards the edge of the array.

\subsection{Example Images and Spectra}
\begin{figure}[h]
\begin{center}
	\subfigure{
	\includegraphics[width=0.3\textwidth]{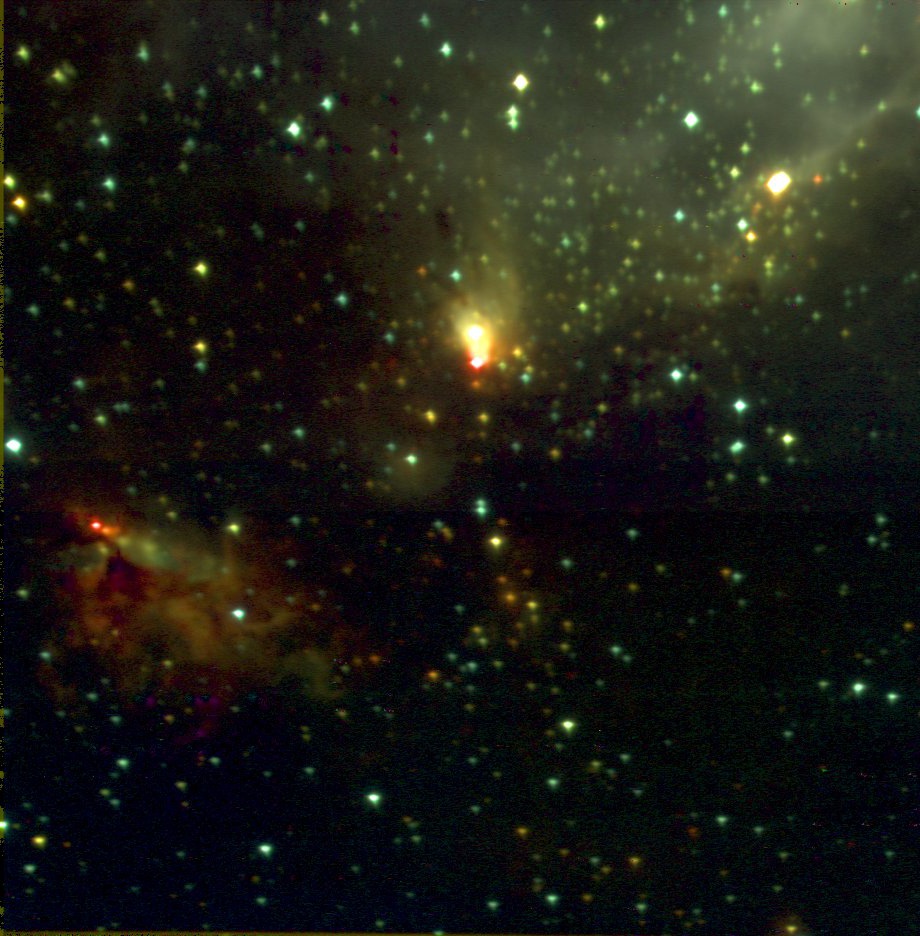} 
	}
	\subfigure{
	\includegraphics[width=0.3\textwidth]{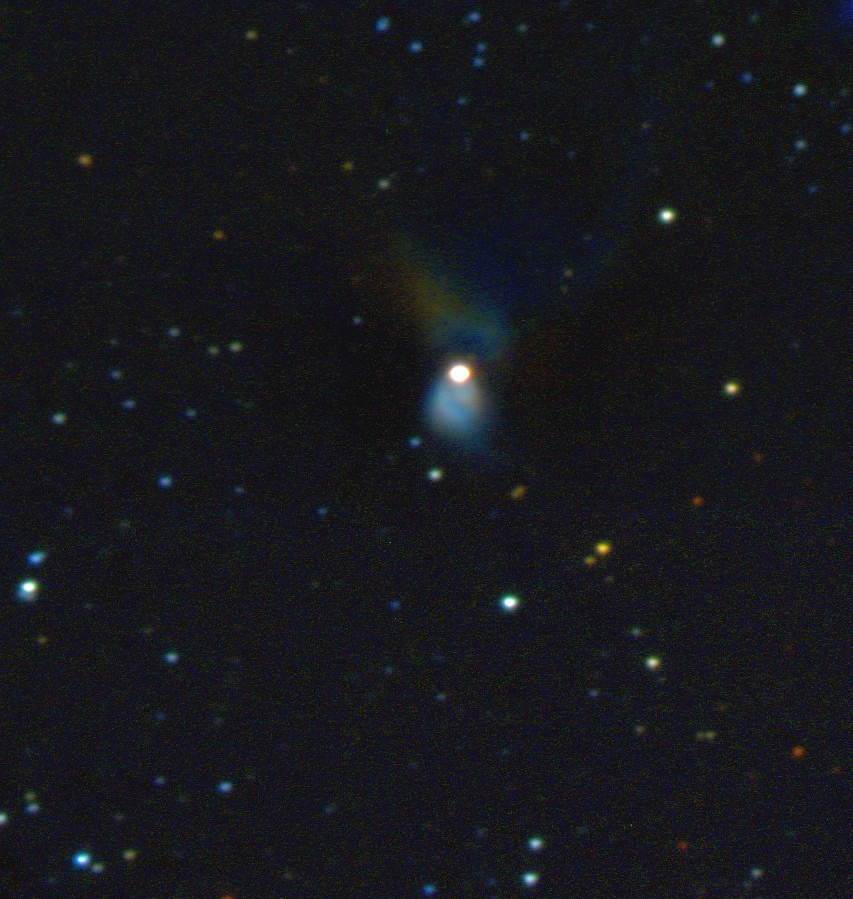} 
    }
    \subfigure{		
	\includegraphics[width=0.3\textwidth]{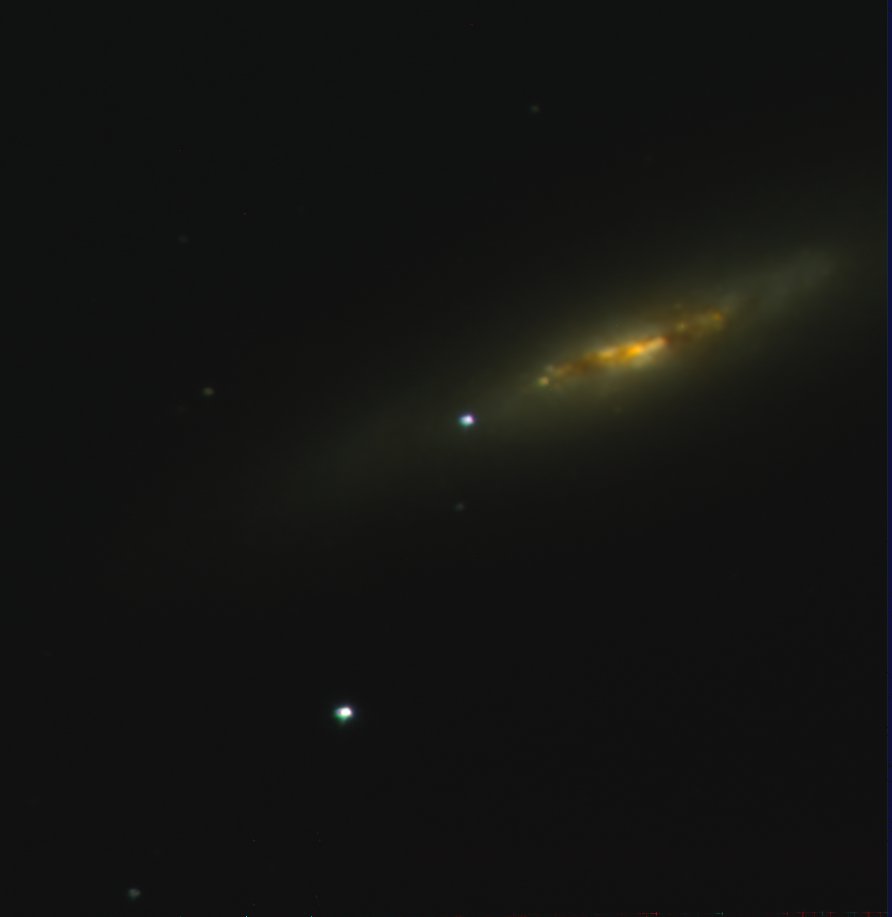} 
	}

\end{center}
\caption{Various RGB color composite images generated using TIRSPEC \textit{J} (blue), \textit{H} (green) and \textit{K$_s$} (red) images. On the left is NGC 7538 (star-forming region), at the middle is V2494 Cyg (FU Ors family of outburst source), and on the right is SN2014J (Supernova in M82).}
\label{fig:GoodRGBimages}
\end{figure}

\begin{figure}[h]
\begin{center}
\subfigure{
\includegraphics[width=0.48\textwidth]{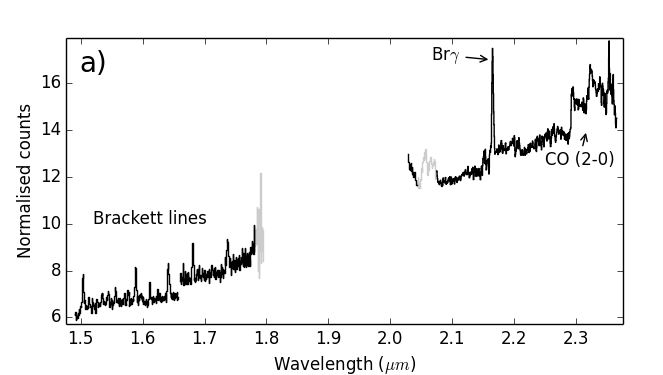} 
}
\subfigure{
\includegraphics[width=0.48\textwidth]{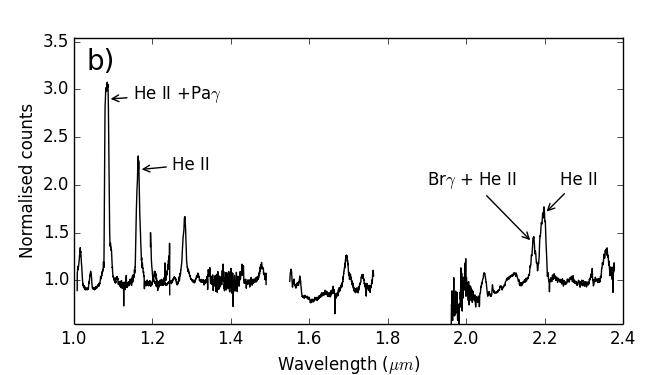} 
}
\subfigure{
\includegraphics[width=0.48\textwidth]{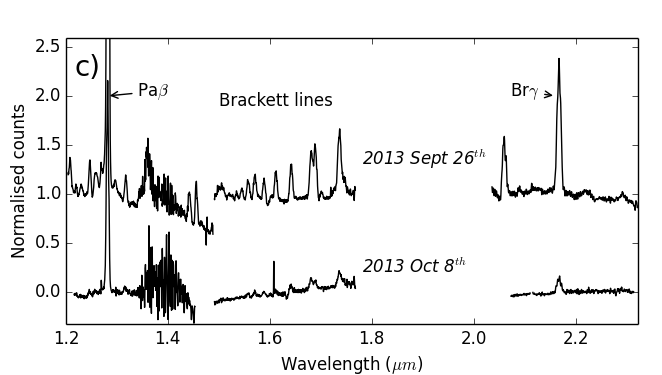} 
}
\subfigure{
\includegraphics[width=0.48\textwidth]{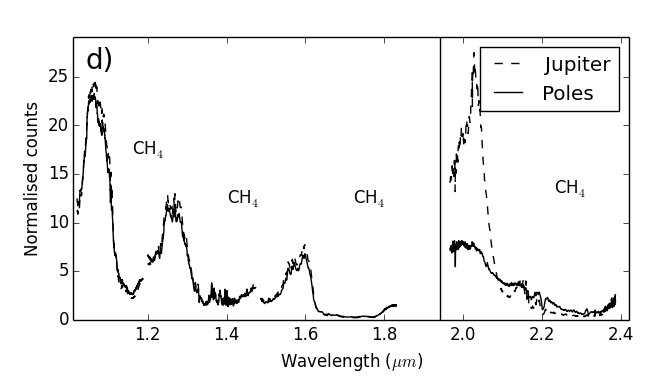} 
}

\subfigure{
\includegraphics[width=0.8\textwidth]{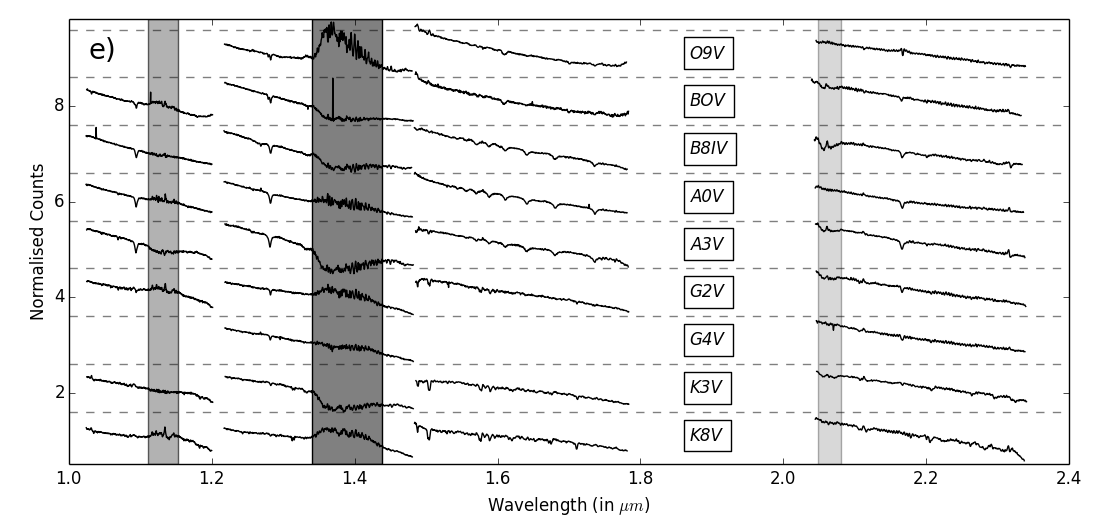} 
}

\end{center}
\caption{Sample of the NIR spectra taken with TIRSPEC. a) PV Cep (FU Ori kind of outburst source with outflow jets), b) WR 6 (Wolf Rayet star), c) Nova Del 2013 (NIR Spectral evolution of Nova), d) Spectrum of Jupiter and its poles (\textit{K} band spectrum is scaled for clarity), e) Stellar Spectra of various main-sequence spectral type stars.}
\label{fig:GoodSpectra}
\end{figure}

During the engineering runs, to test the scientific capabilities of the instrument, we observed various astrophysical sources in imaging as well as in spectroscopy modes. The objects observed range from star-forming regions, transient sources like young low mass star eruptive variables, novae \citep{ninan13b}, supernovae \citep{srivastav14} and Wolf Rayet stars to objects like white dwarfs, our solar system planets, and stars from various positions in the main sequence Hertzsprung-Russell diagram. Figure \ref{fig:GoodRGBimages} shows color composite images created from the \textit{J}, \textit{H} and  \textit{K$_s$} images of some of the sources observed with TIRSPEC. Figure \ref{fig:GoodSpectra} shows the NIR spectra obtained from various kinds of sources taken with TIRSPEC.

\section{Data Reduction Tool}
\label{sec:pipeline}
We have developed the data reduction tool for analysis of TIRSPEC data with a philosophy to automate all the repetitive steps in data reduction, without any compromise on the quality of the reduced data. This semi-automated pipeline requires human guidance only in flagging good and bad frames, choosing sources of interest for photometry etc. This tool is especially useful when many nights of data of a particular source have to be reduced together.
Codes are written in Python, using standard modules used by the  astronomy community like astropy\footnote{\citet{astropy13}}, numpy\footnote{NumPy is the fundamental package for scientific computing with Python \citep{oliphant07}}, matplotlib\footnote{\citet{hunter07}} and PyRAF\footnote{PyRAF is a product of the Space Telescope Science Institute, which is operated by AURA for NASA}. All the codes are written in modular form with inbuilt documentation, so users can easily modify or create their own pipeline by importing the codes as modules. The latest version of codes (under GNU GPLv3 license) are regularly updated on the instrument website\footnote{\url{http://web.tifr.res.in/~daa/tirspec/}}.
The data reduction tool has mainly three parts as explained below.

\subsection{Slope image generation}
The TIRSPEC and the telescope are controlled remotely by the observer from Centre for Research \& Education in Science \& Technology, Hosakote, IIA, Bangalore (India) via a satellite link. In one typical night of observations, the SUTR data acquisition mode in which TIRSPEC is operated, can generate upto 50 GB of raw data. With the existing link bandwidth, it is not possible to download all the raw data to the control room in Bangalore on the same day. Hence, we generate the slope images at the HCT site itself. The first part of the data reduction tool, which does dark subtraction, intelligent pixel masking, and slope fitting (see Section \ref{sec:ReadoutAndImgGen}) is run on the TIRSPEC computer at Hanle. This part is fully automated and run at the end of every observation night. The data that is received by the observer in the Bangalore control room are the dark-subtracted slope images. Two log files are generated and provided to the observer, one containing the log of the image generation pipeline, and other containing a summary of all observations done during the night.
The same code can be imported as a module in an interactive Python terminal if the user wants to work on the raw SUTR data.

\subsection{Photometry}
The photometric part of the TIRSPEC data reduction tool, does basic reduction and contains options to do aperture photometry as well as PSF photometry using \textit{daophot} IRAF binaries. In order to not compromise on the quality of data reduction, selection of good and bad frames is done by human supervision. The code also allows users to run the script in an interactive mode, which allows the user to verify the steps during the process of execution. Photometric magnitude outputs are written to a log file.

\begin{figure}[h]
\begin{center}
\includegraphics[width=0.4\textwidth]{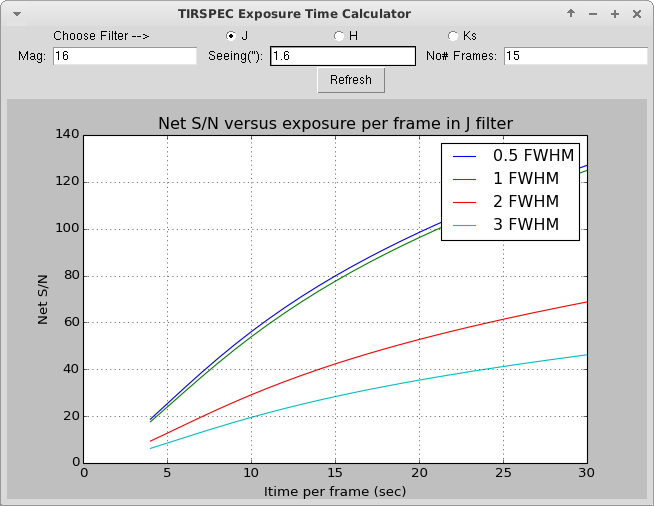} 
\end{center}
\caption{Screenshot of the S/N ratio estimator. Various curves show the S/N ratio that can be obtained for corresponding photometry aperture radius as a function of exposure time per frame.}
\label{fig:ExpCalculatorScreenshot}
\end{figure}

The S/N ratio calculation in the SUTR readout is slightly more complicated than simple double-correlated readouts. We have made an exposure time calculator for TIRSPEC, which first creates a model star profile based on seeing conditions, star magnitude, sky brightness and zero point magnitude of \mbox{TIRSPEC}. The S/N ratio for a set of aperture radii is then numerically calculated using gain, readout noise and number of non-destructive readouts in the exposure using the error propagation formula of the SUTR mode (see \citet{robberto07} for the equations). Figure \ref{fig:ExpCalculatorScreenshot} shows the screenshot of this tool.

\subsection{Spectroscopy}
The spectroscopy part does basic reduction and finally produces 1D wavelength-calibrated spectrum. Human supervision is again needed for identifying the good frames to be reduced. It is also advisable to run this task in an interactive mode to cross check the automated argon line identification during wavelength calibration.
Once the wavelength-calibrated spectra of the science target as well as a standard star are obtained, one can use the Telluric Correction tool to remove the telluric absorption lines in the spectra.

\subsubsection{Telluric Correction Tool}
NIR spectra contain many telluric lines and their removal is done by dividing with the spectrum of a spectroscopic NIR standard star taken within the nearby airmass immediately before or after the target observation. Along with telluric lines, this also removes the instrument response and gives us a continuum-corrected spectra. This step also removes fringes in the \textit{K} band spectra. The Telluric Correction tool is a fully interactive software which removes the standard star's intrinsic stellar lines, align and scale before dividing and finally applies black body temperature correction to obtain the continuum-corrected spectrum of the source.

\section{Conclusion}
This paper discusses the successful installation, characterization and engineering phases of the TIRSPEC. As seen from the results, the TIRSPEC mounted on HCT, Hanle is quite a competitive NIR instrument which has opened up competitive NIR spectroscopy to the Indian astronomy community. 
After successful characterization and performance analysis runs on the telescope, TIRSPEC is now officially released to the astronomy community for science observations from May 1, 2014. It is currently being used heavily for science observations by various astronomers in India. 
In the near future we plan to change the 10" short slits to 50" slits of the same widths. This will allow for dithering of the star position along the slit while taking spectra.
Subarray readout capability of the array will also be implemented soon, which will allow us to take much faster readouts of a small section of the array by compromising on the FoV.

\textit{Note added in proof:} In July 2014, we upgraded all the 10" slits to 50" slits of the same widths.

\section*{Acknowledgments}
We thank referee Dr. Rainer Lenzen for giving us invaluable comments and suggestions that improved the content of the paper. The authors would like to thank the staff at the HCT, Hanle for their co-operation during the installation and characterization and daily maintenance of the TIRSPEC. We especially thank Mr. Angchuk Dorje, Mr. Tsewang Dorjai and Mr. Sonam Jorphail who made crucial contributions during the installation phase.
We would also like to thank the staff at CREST and especially Prof. B. C. Bhatt and Prof. D. K. Sahu, Hosakote for their enthusiasm and help during all the observation runs with TIRSPEC.

\end{document}